\documentstyle{article}

\author{L. S. F. Olavo\\
Departamento de Fisica - Universidade de Brasilia - UnB\\
70910-900 - Brasilia - D.F. - Brazil}
\title{Quantum Mechanics as a Classical Theory XII: \\
Diffraction and Interference
}

\begin{document}

\maketitle
\begin{abstract}
In this paper we will be concerned with the explanation of the interference
and diffraction patterns observed as an outcome of the Young double slit
experiment. We will show that such explanation may be given {\it only} in
terms of a corpuscular theory, which has been our approach since the first
paper of this series. This explanation will be accomplished with an
extension we make here of the domain of applicability of the Born-Sommerfeld
rules that we derived in paper XI of this series.
\end{abstract}

\section{Introduction}

The problems referring to the explanation of the interference and
diffraction patterns one observes in a double slit experiment, for example,
are usually considered as those for which a corpuscular theory seems much
inadequate to deal with.

We have already discussed, very briefly and only qualitatively \cite{eu3},
about the possibility of giving this explanation using only the corpuscular
picture. We will now use the results of paper XI of this series \cite{eu11}
to show quantitatively that this is actually the case---that a corpuscular
explanation for the phenomena of interference and diffraction do exist.

In the second section we will generalize the conclusions of paper \cite{eu11}
with respect to the connections between the Born-Sommerfeld quantization
rules and the formalism based upon the Schr\"odinger equation. This
extension will be needed to include within the domain of applicability of
the Bohr-Sommerfeld rules systems that are not periodic on configuration
space. We will show that these rules are associated with more general
symmetry properties of the physical systems---and include the periodicity as
a particular case.

The third section will deal with the diffraction of particles by a spatially
periodic system. This study has nothing new and was in fact already
developed in the early days of quantum mechanics \cite{Duane}. It will be
discussed here for completeness. The approach we use here is based upon the
Bohr-Sommerfeld rules and its relations with the Schr\"odinger equation
approach were already clarified in paper XI and the previous section.

The fourth section will take into account the problem slightly different of
the diffraction of particles by an aperture. The formalism to be used will
be the same of the third section.

We will, in the fifth section, deal with the problem of interference and
diffraction of particles due to the presence of a double slit Young
interferometer. We will show that our results agree perfectly well with
those found in the literature based upon an undulatory approach.

The last section will be devoted to the conclusions.

\section{Symmetry and Quantization}

In a previous paper \cite{eu11}, we showed that the probability amplitudes
in configuration space may be obtained from the infinitesimal Wigner-Moyal
transformation defined over the phase space as 
\begin{equation}
\label{12.1}\psi (q+\delta q;t)=\int {e^{\frac i\hbar p\delta q}\phi
(q,p;t)dp}.
\end{equation}
In this case, we showed that, if the system has a configuration space
periodicity, given by 
\begin{equation}
\label{12.2}\psi (q+Q;t)=\pm \psi (q;t),
\end{equation}
where $Q$ is the period, then we get the Bohr-Sommerfeld quantization rules 
\begin{equation}
\label{12.3}\oint {p(q)dq}=\left\{ 
\begin{array}{l}
nh
\mbox{ or } \\ (n+1/2)h
\end{array}
\right. ,
\end{equation}
if in (\ref{12.2}) we have the sign $+$ or $-$, respectively. We also showed
there that these rules are valid for each individual system composing the 
{\it ensemble}.

However, one should note that the periodicity argument is not, in any sense,
essential in the derivation. Indeed, even if the amplitude $\psi(q;t)$ is
not periodic, in the sense that equality (\ref{12.2}) is valid for any
integer multiple of the period $Q$, this equality may still be valid for
just one specific value of $Q$. In other words, it means that the value of
the amplitudes at a given point $q$ shall be equal (except for a signal) to
their value at the point $q+Q$, for some $Q$.

In this case, the same quantization rules given by expression (\ref{12.2})
apply and it is in this sense that they will be used in section four and
five.

\section{Laue Diffraction}

We call a Laue diffraction that one in which some particle collides with a
periodic structure (e.g. a crystal lattice) making an angle $\theta $ with
respect to some axis perpendicular to the surface of this crystal and is
elastically reflected by it.

If we suppose that the crystal has identical planes spaced regularly by a
constant distance $d$ (see figure 1) measured with respect to the z-axis, we
may conclude \cite{Pauling} that the quantization rules in the z-direction
becomes 
\begin{equation}
\label{12.4} \oint{p_z dz}=\int_{0}^{d}{p_z dz}=n_z h \mbox{ or } p_z=\frac{%
n_z h}{d}, 
\end{equation}
where we used that the crystal is a periodic structure of period $Q=d$ in
this direction.

Any interaction of the crystal with an incident particle must be such as to
let the crystal momentum $p_z$ quantized, varying it by the amount 
\begin{equation}
\label{12.4.1}\Delta p_z=\frac{(\Delta n_z)h}d=\frac{nh}d.
\end{equation}
Looking at figure 1 it is easy to see that this interaction induces a
momentum transfer from the incident particle to the crystal (or {\it vice
versa}) given by 
\begin{equation}
\label{12.5}p_z=\frac{2h}\lambda \sin (\theta ),
\end{equation}
where we wrote 
\begin{equation}
\label{12.6}p=\frac h\lambda ,
\end{equation}
being $\lambda $ a characteristic length%
\footnote{Note that the introduction
of this variable $\lambda$ is inessential to the problem. We make this
deffinition as to make the final formulae easier to compare with those
found in the literature.} related to the particle. In this case, the
equation for the momentum balance gives the expression 
\begin{equation}
\label{12.7}n\lambda =2d\sin (\theta ),
\end{equation}
which is the equation for the maximum intensity positions of the Laue (or
Bragg) diffraction of particles by a periodic structure.

This allows us to say that the diffraction pattern one obtains making a flux
of particles to hit upon the surface of a crystal lattice is exactly the
same one gets using the undulatory approach%
\footnote{Except, of course, for
the intensities, since they are not present in the Bohr-Sommerfeld
treatment.}, precluding one to make any reference to the objective existence
of ``material waves'' or, which is the same, to the concept of duality;
being necessary only to consider the quantization of momentum transfer.

\section{Diffraction by an Aperture}

In this problem we do not have in general any periodic structure. We have,
instead, as shown in figure 2, a flux of particles incident at right angles
upon a screen on which we made an aperture of size $a$, and being scattered
by the atoms on the borders of this aperture at angles $\theta $.

The important thing to stress here is that this system has a spatial
symmetry. Indeed, it is easy to note that the probability amplitude shall
have the same value (except for a sign) in each border of the aperture. This
gives 
\begin{equation}
\label{12.8}\psi (+a/2)=\pm \psi (-a/2),
\end{equation}
since the probability density is the same at both points.

Using the results of the second section of this paper we may write, using
equation (\ref{12.8}), 
\begin{equation}
\label{12.9} \int_{-a/2}^{+a/2}{p_z dz}=n h, 
\end{equation}
where all the quantities have the same meaning as in the previous section%
\footnote{Note that the integral is being not taken on a closed trajectory,
since the system does not have periodicity.}.

Making all the calculations we get 
\begin{equation}
\label{12.10}n\lambda =a\sin (\theta ),
\end{equation}
which is the same expression for the intensity minima of the diffraction
pattern%
\footnote{The present formalism is not able to say whether we are
at the minima or at the maxima of the intensities for the very reason that
the intensities are not one of its scope.}.

We must stress here the distinctions between the explanation of the two
diffraction behavior we have analyzed. The Laue diffraction is due to a
spatial periodicity possessed by the crystal lattice. The diffraction by an
aperture is associated with a particular symmetry of the system (a rotation
of $\pi $ about the axis perpendicular to the screen), being unnecessary
that the screen itself be periodic. This is the explanation for the
appearance, in the first case, of the crystal period, while in the second
case we have the appearance of the aperture dimension.

We may say, following the interpretation of the last section, that the
introduction of the aperture in the screen changes or redefines the relation
of momentum transfer between the screen and the incident particles. These
changes are responsible for the quantized scattering relations given by 
(\ref{12.10}).

\section{Double Slit}

We may account for this problem in exactly the same manner as we did with
the diffraction by an aperture in the last section.

To begin with, consider first figure 3. In this figure we have two apertures
of length $a$, made over some screen, symmetrically placed as related with 
the $x$-axis. These apertures are at heights $z=+c/2$ and $c=-c/2$ of the $x$%
-axis. A flux of particles passes through these apertures and are
deflected as they interact with their borders.

Such a system possesses, as becomes clear in figure 3, the following
symmetries 
\begin{equation}
\label{12.11} \psi(-c/2-a/2)=\psi(+c/2+a/2) \mbox{ and } \psi(-c/2)=%
\psi(+c/2). 
\end{equation}
The last one of these symmetries gives the momentum quantization 
\begin{equation}
\label{12.12} p_z c=n_1 h, 
\end{equation}
where $n_1$ is some integer number.

The first symmetry in (\ref{12.11}) gives 
\begin{equation}
\label{12.13} p_z(c+a)=(n_2+1/2) h, 
\end{equation}
where $n_2$ is a second integer number%
\footnote{The criterion for using
the second, and not the first, possibility in (\ref{12.3}) to represent
the intensity maxima, is certainly arbitrary within this theory, as we
have already stressed. The Bohr-Sommerfeld rules give only the positions
of the maxima an minima but not a means of distinguishing them both.}.

Solving the equations (\ref{12.12}) and (\ref{12.13}) we get 
\begin{equation}
\label{12.14} \left\{ 
\begin{array}{l}
n\lambda=c\sin(\theta); \\ 
(m+1/2)\lambda=a\sin(\theta), 
\end{array}
\right. , 
\end{equation}
where we made $n=n_1$ and $m=n_2-n_1$.

The first expression reflects the intensity maxima condition related with
the interference pattern for a double slit spaced by a distance $c$, while
the second expression reflects the conditions, also for the intensity
maxima, related with the diffraction by an aperture of width $a$.

Except for the information about the relative intensities, as we already
stressed, these conditions for the intensities extrema are exactly those
obtained in the literature, when we use the undulatory approach to the
problem. In this case, as was presented in the previous sections, it was not
necessary to go beyond the scope of a corpuscular theory to give a
mathematically grounded explanation of the phenomena considered.

In the same sense as with the diffraction by an aperture problem, we may say
that the introduction of a second aperture on the screen modifies the
quantized momentum transfer relations between the particle and the screen
itself. These new relations will be responsible for the interference and
diffraction patterns%
\footnote{We have been using the words interference
and diffraction throughout this paper. This is justified by the wide
use this nomenclature has in the literature. Of course, our approach
denies the strict use of these words.}.

\section{Conclusion}

As we have said in the introductory section, the problems of diffraction and
interference---mainly the later---were always considered as the ones
forbidding a phenomenological interpretation of the quantum formalism based
only upon a corpuscular model.

This paper has shown that such an interpretation is indeed possible,
avoiding the need to appeal to an undulatory description. Each individual
particle is scattered at quantized angles. After more and more particles are
scattered (more systems composing the {\it ensemble} are considered), we
begin to see the figures, known as the interference and/or diffraction
patterns. These patterns are, of course, a property of the {\it ensemble},
while each scattering refers to only one particle. That's why the amplitudes
refers to {\it ensembles} while the Bohr-Sommerfeld rules refer to
individual systems.

The behavior of this group of systems (of the {\it ensemble}) will be of an
undulatory character which is described by the Schr\"odinger equation by
means of the amplitudes. Since this undulatory behavior refers to the {\it %
ensemble} and not to the individual systems, they have no essential
(objective) character. This essential character has to be attributed to the
individual particles, according to what we have done in the postulates of
this theory \cite{eu1}-\cite{eu11}.


\newpage

\vspace{20pt}

\unitlength=1.00mm \special{em:linewidth 1pt} \linethickness{1pt}

\begin{figure}
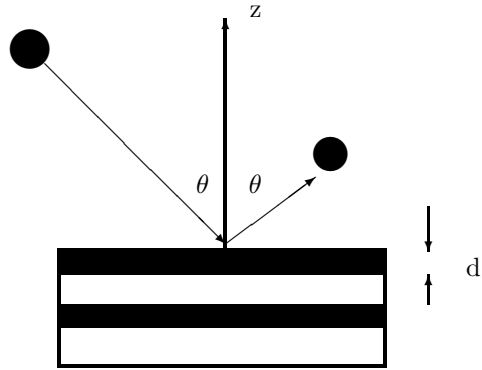
(76.00,57.00)
\put(21.00,10.00){\framebox(43.00,15.00)[cc]{}}
\put(70.00,31.00){\vector(0,-1){6.00}}
\put(70.00,18.00){\vector(0,1){4.00}}
\put(76.00,23.00){\makebox(0,0)[cc]{d}}
\put(47.00,57.00){\makebox(0,0)[cc]{z}}
\put(57.00,38.00){\circle*{4.47}}
\put(17.00,52.00){\circle*{5.20}}
\put(40.00,34.00){\makebox(0,0)[cc]{$\theta$}}
\put(47.00,34.00){\makebox(0,0)[cc]{$\theta$}}
\put(43.00,25.00){\vector(0,1){31.00}}
\put(19.00,50.00){\vector(1,-1){24.00}}
\put(43.00,26.00){\vector(0,1){0.00}}
\put(43.00,26.00){\vector(4,3){12.00}}
\put(21.00,22.00){\rule{43.00\unitlength}{3.00\unitlength}}
\put(21.00,15.00){\rule{43.00\unitlength}{3.00\unitlength}}
\caption{Laue (or Bragg) diffraction by a periodic structure.}
\end{figure}

\newpage

\vspace{20pt}

\begin{figure}
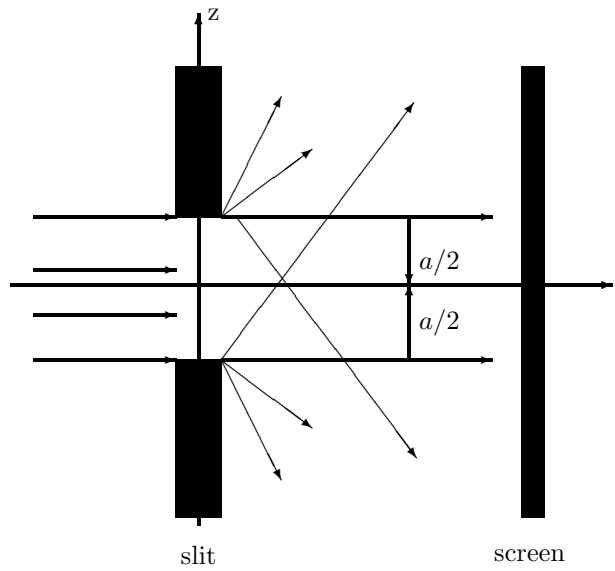
(83.00,77.01)
\put(25.00,10.00){\rule{6.00\unitlength}{21.00\unitlength}}
\put(25.00,50.00){\rule{6.00\unitlength}{20.00\unitlength}}
\put(6.00,31.00){\vector(1,0){19.00}}
\put(6.00,37.00){\vector(1,0){19.00}}
\put(6.00,43.00){\vector(1,0){19.00}}
\put(6.00,50.00){\vector(1,0){19.00}}
\put(31.00,50.00){\vector(1,2){8.00}}
\put(31.00,50.00){\vector(4,3){12.10}}
\put(31.00,31.00){\vector(4,-3){12.10}}
\put(31.00,31.00){\vector(1,-2){8.00}}
\put(56.00,41.00){\vector(0,1){0.03}}
\put(56.00,41.00){\vector(0,-1){0.03}}
\put(71.00,10.00){\rule{3.00\unitlength}{60.00\unitlength}}
\put(60.00,44.00){\makebox(0,0)[cc]{$a/2$}}
\put(60.00,36.00){\makebox(0,0)[cc]{$a/2$}}
\put(81.00,41.00){\vector(1,0){1.96}}
\put(28.00,77.00){\vector(0,1){0.01}}
\put(72.00,5.00){\makebox(0,0)[cc]{screen}}
\put(28.00,5.00){\makebox(0,0)[cc]{slit}}
\put(30.00,77.00){\makebox(0,0)[cc]{z}}
\put(28.00,9.00){\vector(0,1){68.00}}
\put(3.00,41.00){\vector(1,0){80.00}}
\put(56.00,50.00){\vector(0,-1){9.00}}
\put(56.00,31.00){\vector(0,1){10.00}}
\put(31.00,50.00){\vector(1,0){36.00}}
\put(31.00,31.00){\vector(1,0){36.00}}
\put(33.00,50.00){\vector(3,-4){24.00}}
\put(31.00,31.00){\vector(3,4){25.67}}
\caption{Diffraction through an apperture.}
\end{figure}

\newpage

\vspace{20pt}

\begin{figure}
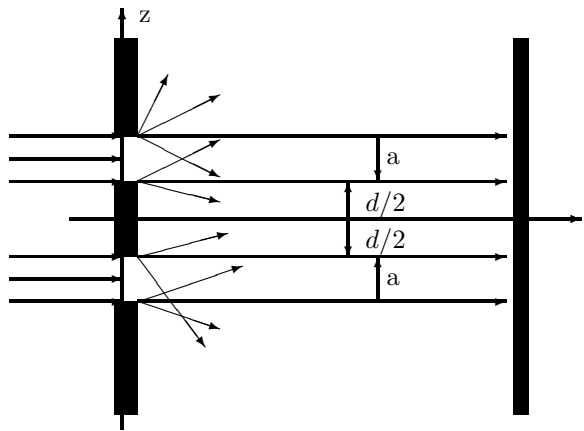
(81.02,59.04)
\put(19.00,5.00){\rule{3.00\unitlength}{15.00\unitlength}}
\put(19.00,26.00){\rule{3.00\unitlength}{10.00\unitlength}}
\put(19.00,42.00){\rule{3.00\unitlength}{13.00\unitlength}}
\put(5.00,20.00){\vector(1,0){15.04}}
\put(5.00,26.00){\vector(1,0){15.04}}
\put(5.00,23.00){\vector(1,0){15.04}}
\put(5.00,36.00){\vector(1,0){15.04}}
\put(5.00,39.00){\vector(1,0){15.04}}
\put(5.00,42.00){\vector(1,0){15.04}}
\put(22.00,42.00){\vector(1,2){4.07}}
\put(22.00,42.00){\vector(2,1){10.97}}
\put(22.00,42.00){\vector(2,-1){10.97}}
\put(22.00,36.00){\vector(2,1){10.97}}
\put(22.00,36.00){\vector(4,-1){10.97}}
\put(22.00,26.00){\vector(4,1){12.05}}
\put(22.00,26.00){\vector(3,-4){9.03}}
\put(22.00,20.00){\vector(3,1){13.99}}
\put(22.00,20.00){\vector(3,-1){10.97}}
\put(72.00,5.00){\rule{2.00\unitlength}{50.00\unitlength}}
\put(20.00,59.00){\vector(0,1){0.04}}
\put(78.00,31.00){\vector(1,0){3.02}}
\put(23.00,58.00){\makebox(0,0)[cc]{z}}
\put(13.00,31.00){\vector(1,0){68.00}}
\put(20.00,3.00){\vector(0,1){56.00}}
\put(22.00,42.00){\vector(1,0){49.00}}
\put(22.00,36.00){\vector(1,0){49.00}}
\put(22.00,26.00){\vector(1,0){49.00}}
\put(22.00,20.00){\vector(1,0){49.00}}
\put(50.00,31.00){\vector(0,-1){5.00}}
\put(50.00,31.00){\vector(0,1){5.00}}
\put(54.00,42.00){\vector(0,-1){6.00}}
\put(54.00,20.00){\vector(0,1){6.00}}
\put(56.00,23.00){\makebox(0,0)[cc]{a}}
\put(56.00,39.00){\makebox(0,0)[cc]{a}}
\put(55.00,33.00){\makebox(0,0)[cc]{$d/2$}}
\put(55.00,28.00){\makebox(0,0)[cc]{$d/2$}}
\caption{Interference and diffraction from a double slit.}
\end{figure}

\end{document}